\documentclass[%
twocolumn,
amsmath,amssymb,
aps,
physrev,
]{revtex4-2}

\usepackage{soul}
\usepackage{graphicx}
\usepackage{dcolumn}
\usepackage{bm}
\usepackage{xcolor}

\begin{document}

\title{\textbf{Localization–Delocalization Transition in Diffusion with Adaptive Resetting} 
}%

\author{Tommer D. Keidar}
\email{Contact author: tommerdavidk@mail.tau.ac.il}
\affiliation{%
School of Chemistry, The Center for Computational Molecular and Materials Science, The Center for Physics and Chemistry of Living Systems, Tel Aviv University, Tel Aviv, Israel
}%
 
\author{Shlomi Reuveni}%
 \email{Contact author: shlomire@tauex.tau.ac.il}
\affiliation{%
School of Chemistry, The Center for Computational Molecular and Materials Science, The Center for Physics and Chemistry of Living Systems, Tel Aviv University, Tel Aviv, Israel
}%

\date{\today}

\begin{abstract}
Stochastic resetting can localize diffusion and generate nonequilibrium steady states, but the conditions under which spatially dependent resetting produces localization remain unclear. Here, we establish a general classification for diffusion under adaptive resetting, where the resetting rate $r(x)$ depends on position. For rates with the asymptotic scaling $r(x)\sim |x|^\lambda$, we identify a sharp threshold at $\lambda=-2$. For $\lambda>-2$, the steady state is localized and exhibits stretched-exponential tails, whereas for $\lambda<-2$, resetting is asymptotically too weak to localize the particle. Precisely at the marginal scaling $r(x)\sim |x|^{-2}$, a qualitatively new regime emerges: the steady state develops power-law tails with a temperature-dependent exponent and exhibits a finite-temperature delocalization transition. Thus, inverse-square resetting plays the role of the logarithmic potential in equilibrium, establishing a nonequilibrium counterpart of the temperature-driven delocalization seen there.
\end{abstract}

\maketitle
Consider a particle with a diffusion coefficient $D=\mu k_BT$, where $\mu$ is the particle mobility, $k_B$ is the Boltzmann constant, and $T$ is the temperature. The particle is diffusing under a potential $U(x)$ where $|U(x)|<\infty$. Will the system reach a steady state, and what will it look like?

If the potential has the asymptotic form $U(x)\simeq\gamma |x|^\lambda$ at large $|x|$, where both $\gamma$ and $\lambda$ are positive, the distribution tails will obey the stretched exponential law $p(x)\sim e^{-\frac{\gamma}{k_BT}|x|^\lambda}$, as dictated by Boltzmann. Moreover, since a stretched exponential is normalizable for any value of ($\gamma/k_BT)$, an equilibrium steady state will be reached at any temperature. 

Potentials that are asymptotically flat, or decreasing, lead to a non-normalizable Boltzmann distribution and thus cannot sustain a steady state. A potential whose asymptotic  growth is logarithmic, i.e., $U(x)\simeq U_0\ln(|x|)$ at large $|x|$, gives power-law tails
$p(x)\sim |x|^{-\frac{U_0}{k_BT}}$. If the energy scale of the potential is larger than the thermal energy, $U_0>k_BT$, this distribution is normalizable, and the system will reach a steady state. Conversely, for $U_0<k_BT$, the system will not reach a steady state, revealing a delocalization transition at a critical temperature $T_c=U_0/k_B$. This kind of transition appears in various systems, including the transition fro normal to anomalous diffusion in cold atoms in optical lattices \cite{Marksteiner1996,lutz2004power,Kessler2010}, the Manning transition in polyelectrolyte solutions \cite{Manning1969}, and the thermal unbinding of logarithmically interacting vortex–antivortex pairs in the Berezinskii–Kosterlitz–Thouless transition \cite{Kosterlitz_1973}.

The equilibrium picture raises a broader question: can one similarly determine whether a steady state exists—and characterize its asymptotic tails—when localization is generated by an intrinsically nonequilibrium mechanism? A paradigmatic mechanism for creating non-equilibrium steady states (NESS) is stochastic resetting. There, a particle's location is reset to the origin at random times \cite{Evans2011PRL, Evans_2013, Eule_2016, Evans_2020, Gupta2022,kundu2024}. For free diffusion, a steady state will emerge as long as the time between resetting events has a finite mean \cite{Nagar2016, Evans_2020}. Stochastic resetting of diffusing particles has been realized experimentally, establishing it as an accessible model system for nonequilibrium statistical mechanics \cite{tal_Friedman2020, besga2020, faisant2021, altshuler2024environmental,vatash2025,biroli_2026Experimental}. 

\textit{Adaptive resetting} is a form of resetting where the resetting rate itself is allowed to depend on the state of the process, e.g., the particle's location \cite{Keidar2025}. In recent years, there has been a growing interest in adaptive resetting protocols \cite{Evans2011JPhysA, Roldan2016, Roldan2017, Plata2020, Tucci2020, Pinsky2020, DeBruyne2020, Ye_2022, Ali_2022, Cantisan2024, Munoz2025, Church2025, TalFriedman2025, Biroli_2026_first, Aspman2026}. For diffusion under adaptive resetting, it was shown that a steady state will emerge if the mean time between resetting events is finite \cite{Keidar2025, Roldan2017}. This is analogous to standard stochastic resetting. However, in adaptive resetting, the statistics of the resetting times are coupled to the underlying process, which is itself stochastic. Specifically, the random time $R$ between resetting events is distributed as
\begin{equation}\label{eq: R distribution}
    \Pr(R< t)=1-\mathbb{E}\left[e^{-\int_0^t r\left(X_s,s\right)ds}\right],
\end{equation}
where $r\left(X_s,s\right)$ is the resetting rate (can depend on state and time), and the expectation is taken over random trajectories $\{X_s,\,0\leq s<t\}$ of the underlying process. Thus, unlike standard stochastic resetting, the resetting-time statistics—and hence the existence of a steady state—are determined jointly by the resetting protocol and the underlying stochastic dynamics. 

\begin{figure*}[t]
    \centering
    \includegraphics[width=0.95\linewidth]{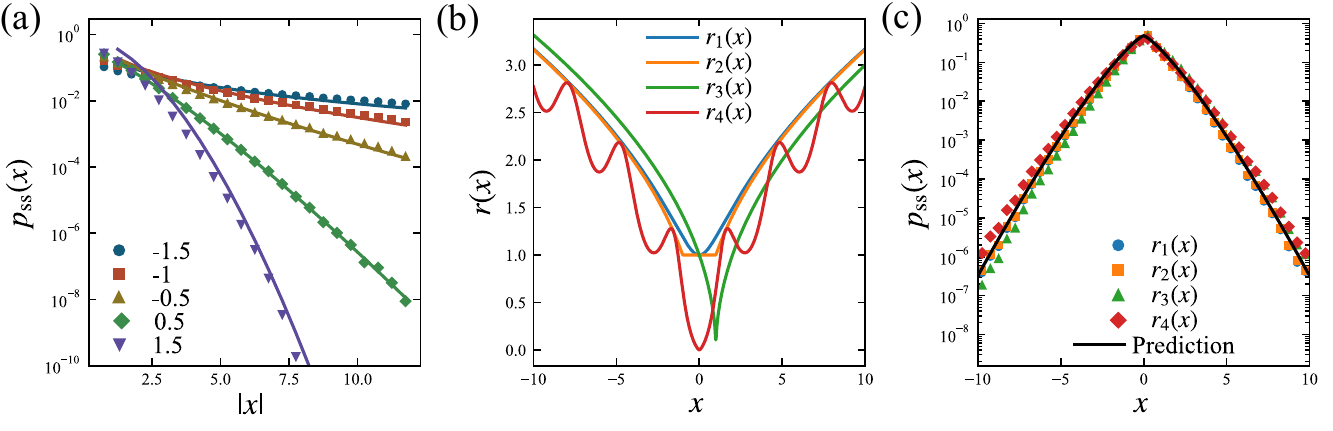}
    \caption{\textbf{a.} The NESS of diffusion ($D=1$) under adaptive resetting with the power-law asymptotics of Eq. (\ref{eq:power_law_rate}). Symbols are numerical evaluations, and solid lines are the theoretical predictions of Eqs. (\ref{eq: stretched exponential}) and  (\ref{eq:streched_exp_paramters}). The resetting rates used are $r(x)=|x|^\lambda$ for $\lambda>0$ and $r(x)=(0.5+|x|)^\lambda$ for $\lambda<0$, with $\lambda\in\{-1.5,\,-1,\,0.5,\,1.5\}$. \textbf{b.} Four different resetting rates with the same power-law asymptotics of $r(x)\sim|x|^{0.5}$. The resetting rates are: $r_1(x)=(1+x^2)^{1/4}$, $r_2(x)=\Theta(1-|x|)+\sqrt{x}\Theta(|x|-1)$ where $\Theta(\cdot)$ is Heavyside step function, $r_3(x)=\sqrt{|x-1|}$, and $r_4(x)=|x|^{1.5}/(|x|+2\cos^2(x))$. \textbf{c.} The NESS for diffusion under each of the resetting rates presented in panel b (symbols), with the asymptotic prediction given by Eqs. (\ref{eq: stretched exponential}) and (\ref{eq:streched_exp_paramters}) (solid line).}
    \label{fig:lambda_great_than_-2}
\end{figure*}
It was hypothesized that if the adaptive resetting rate scales as $\sim|x|^\lambda$ for large $|x|$, the tails of the NESS distribution of diffusion will scale as a stretched exponential, with an exponent of $\lambda/2+1$ \cite{Keidar2025}. This was shown analytically for $\lambda=0,2$ \cite{Evans2011PRL, Roldan2017}, and numerically for $\lambda=1,3$ \cite{Keidar2025}. In this Letter, we establish the result for $\lambda>-2$ and determine both the shape and scale of the resulting stretched-exponential tails. More importantly, we uncover a sharp change in behavior at $\lambda=-2$. At this marginal scaling, the NESS develops power-law tails and exists only below a critical temperature, giving rise to a temperature-driven delocalization transition. For $\lambda<-2$, resetting becomes asymptotically too weak to sustain a steady state. 

These results identify $r(x)\sim |x|^{-2}$ as the resetting analogue of logarithmic confinement and establish a nonequilibrium boundary between localized and delocalized diffusion. In higher dimensions, the stretched-exponential regime remains unchanged, whereas dimensionality modifies the power-law tails at $\lambda=-2$ and lowers the critical temperature. All results are corroborated with extensive numerical analysis, made possible by the efficient numerical method for obtaining properties of processes with adaptive resetting presented in \cite{Keidar2025}.

\textit{Diffusion under power-law adaptive resetting.} The NESS under adaptive resetting protocol $r(x)$ is the solution of the following ordinary-differential equation
\begin{equation}\label{eq: NESS ODE generic}
        D\frac{d^2p(x)}{dx^2}-r(x)p(x)+Q\delta(x)=0
\end{equation}
where $D$ is the diffusion coefficient, and $Q=\int_{-\infty}^\infty r(x)p(x)\,dx$. The second term on the left-hand side accounts for the probability loss due to resetting, and the third is the probability gain at the origin due to resetting. At the steady-state, the sum of all three contributions must equate to zero. For the solution to describe a physical probability distribution, it must be positive and normalizable. We will consider cases where for $-\infty<x<\infty,\,$ the resetting rate is finite $0\leq r(x)<\infty$, and has power-law asymptotics
\begin{equation}\label{eq:power_law_rate}
    r(x)\simeq r_0|x|^\lambda,
\end{equation}
for $|x|\gg 1$. Here, $\lambda\in\mathbb{R}$, and $r_0>0$ is a constant with dimensions $[\text{Time}^{-1}\cdot\text{Length}^{-\lambda}]$. 

To obtain the tails of the distribution, we can plug this asymptotic behaviour into Eq. (\ref{eq: NESS ODE generic}), resulting in
\begin{equation}\label{eq: NESS ODE tails}
    \frac{d^2p(x)}{dx^2}-\frac{r_0}{D}x^\lambda p(x)=0,
\end{equation}
where, without loss of generality, we took $x>0$. Next, we use the ansatz, hypothesized in \cite{Keidar2025}: the NESS tails have a stretched exponential form
\begin{equation}\label{eq: stretched exponential}
    p(x)\propto e^{-b|x|^\alpha},
\end{equation}
where $b,\,\alpha>0$. Plugging this ansatz into Eq. (\ref{eq: NESS ODE tails}), results in
\begin{equation}
    \begin{cases}
        \alpha bx^{\alpha-2}(\alpha bx^{\alpha}-\alpha+1)-\frac{r_0}{D}x^\lambda=0&\alpha\neq1,\\
        b^2-\frac{r_0}{D}x^\lambda=0&\alpha=1.
    \end{cases}
\end{equation}
As we are looking for an asymptotic solution, and because $\alpha>0$, we have $\alpha b x^\alpha\gg 1-\alpha$. We thus get that $\alpha$ and $b$ are set by $\alpha^2b^2x^{2(\alpha-1)}=\frac{r_0}{D}x^\lambda$, whose solution is 
\begin{equation}\label{eq:streched_exp_paramters}
    \alpha=\frac{\lambda}{2}+1,\quad b=\frac{2}{\lambda+2}\sqrt{\frac{r_0}{D}}.
\end{equation}
The values predicted for $\alpha$ by Eq. (\ref{eq:streched_exp_paramters}) agree with the full analytical solutions that were obtained for $\lambda=0$ \cite{Evans2011PRL} (constant rate resetting, Laplace tails) and $\lambda=2$ \cite{Roldan2017} (parabolic resetting, Gaussian tails), and the numerical results for $\lambda=1,3$ \cite{Keidar2025}.

In Fig. \ref{fig:lambda_great_than_-2}(a), we compared this result to extensive numerical evaluations of the NESS under adaptive resetting rates that obey Eq. (\ref{eq:power_law_rate}). 
To test how the asymptotic solution fits moderate $x$ values, four different functions that share the same power-law asymptotic behavior, $r(x)\sim\sqrt{|x|}$, were used (Fig. \ref{fig:lambda_great_than_-2}(b)). The resulting NESS are presented in Fig. \ref{fig:lambda_great_than_-2}(c). Although Eqs. (\ref{eq: stretched exponential}) and (\ref{eq:streched_exp_paramters}) were derived by taking the large $|x|$ limit, they provide good approximations even for moderate $x$-values.  Note that predictions for the tails hold up to a prefactor in Eq. (\ref{eq: stretched exponential}). This prefactor is not universal, as it depends on the bulk behaviour of the NESS through the normalization constraint. To account for this, we chose the prefactor such that the 
prediction in Eq. (\ref{eq: stretched exponential}) agrees with simulations at $|x|=12$. Finally, in the End Matter, we further corroborate the asymptotic result by plotting the NESS in Fig. \ref{fig:lambda_great_than_-2} (a, c) with a transformed x-axis $b|x|^\alpha$ and showing that all curves have a slope of $-1$ when the y-axis is on a log scale, as predicted by Eqs. (\ref{eq: stretched exponential}) and (\ref{eq:streched_exp_paramters}).

To evaluate the NESS in Figs. \ref{fig:lambda_great_than_-2}(a, c), we used the procedure described in \cite{Keidar2025}. First, we simulated an ensemble of $10^4$ diffusive trajectories, each consisting of $N=12\cdot10^3$ time steps, with a time step length of $\Delta t=0.05$. For each trajectory and resetting-rate function $r(x)$, we then calculated the probability that the trajectory survives without resetting up to step $n$: $\approx e^{-\sum_{i=1}^{n-1}r(x_i)\Delta t}$, for all $1\leq n \leq N$ and where $x_i$ is the position at the $i$-th time step. To estimate the NESS, we weigh the probability of finding a diffusive trajectory at a given point by the probability of reaching this point without being reset along the way. Because this is proportional to the fraction of time a Brownian motion will spend in this location conditioned on not being reset, this gives the NESS up to normalization.

\textit{Diffusion under $r(x)\sim r_0/x^2$ adaptive resetting.} Normalizability of the stretched-exponential solution in Eqs. (\ref{eq: stretched exponential}) and (\ref{eq:streched_exp_paramters}) requires $\alpha,b>0$, restricting it to $\lambda>-2$. Next, we show that in the marginal case $\lambda=-2$, the NESS instead develops power-law tails with a temperature-dependent exponent. As this exponent approaches unity from above, the distribution becomes non-normalizable, giving rise to a delocalization transition.

For $\lambda=-2$, Eq. (\ref{eq: NESS ODE tails}) becomes
\begin{equation}\label{eq: NESS tails Lorentz}
    \frac{d^2p(x)}{dx^2}-\frac{r_0}{D}\frac{1}{x^2}p(x)=0.
\end{equation}
Interestingly, in this case, $r_0/D$ is dimensionless, suggesting self-similar behavior. 
We  therefore use a power-law ansatz for the tails of the distribution
\begin{equation}\label{eq:power_law}
    p(x)\propto |x|^{-\eta}.
\end{equation}
For it to be normalizable, we must have $\eta>1$. Plugging this ansatz into Eq. (\ref{eq: NESS tails Lorentz}) results in $\eta(1+\eta)x^{-(\eta+2)}=\frac{r_0}{D}x^{-(\eta+2)}\Rightarrow\eta^2+\eta-\frac{r_0}{D}=0$.
Solving this quadratic equation for $\eta$ gives the positive solution
\begin{equation}\label{eq: power law NESS power}
    \eta=\frac{1}{2}\left(\sqrt{1+\frac{4r_0}{D}}-1\right)=\frac{1}{2}\left(\sqrt{1+\frac{4r_0}{\mu k_BT}}-1\right),
\end{equation}
where in the second equality we have invoked the Einstein-Smoluchowski relation. We thus find a NESS with power-law tails and a temperature-dependent power-law exponent. This means that there is a series of phase transitions as $\eta$ crosses integer values; above each critical temperature, an additional moment of the steady-state distribution becomes ill-defined. For example, for $T>r_0/(12\mu k_B)$ the steady-state variance diverges.

\begin{figure}[t]
    \centering
 \includegraphics[width=1\linewidth]{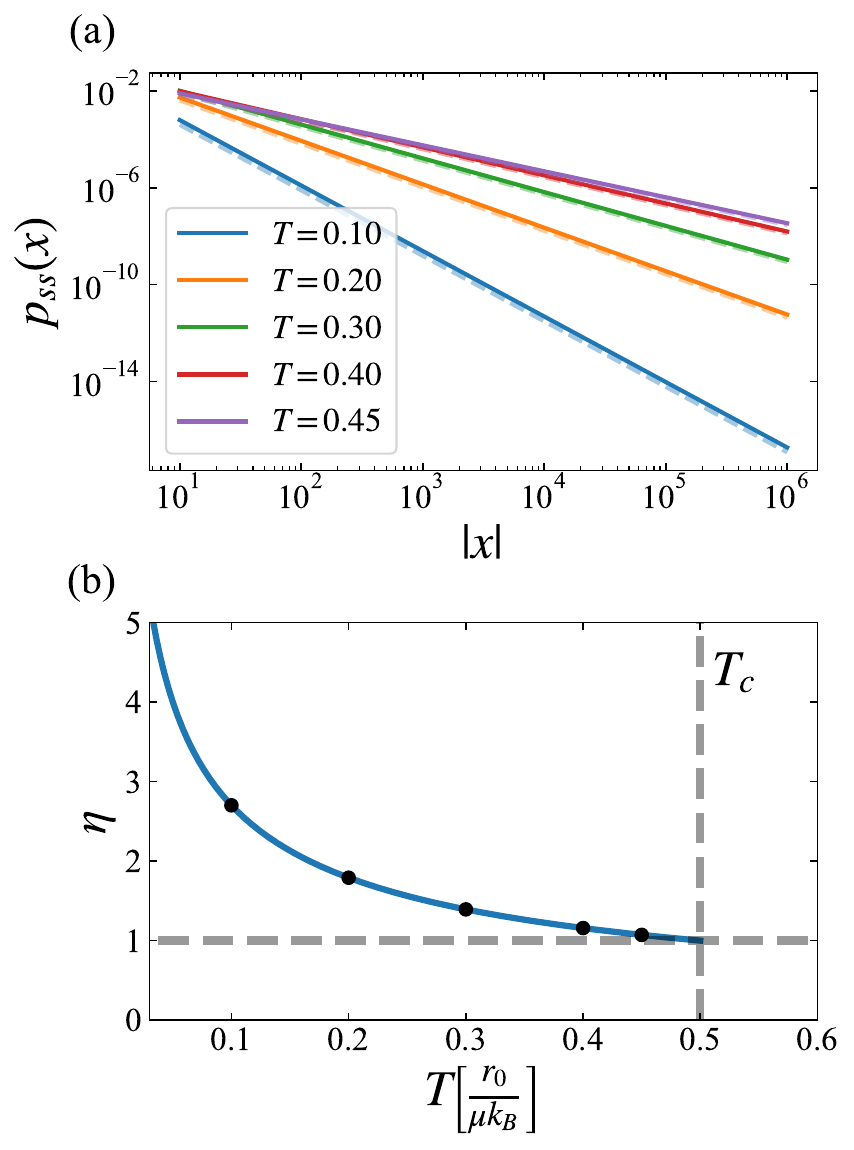}
    \caption{(a) The NESS solution of Eq. (\ref{eq: NESS ODE generic}) with $r(x)=(1+x^2)^{-1}$ for different temperatures. The solid line comes from a numerical solution of the equation, and dashed lines are given by Eqs. (\ref{eq:power_law}) and (\ref{eq: power law NESS power}). Temperature is measured in $r_0/(\mu k_B)$. (b) Plot of Eq. (\ref{eq: power law NESS power}). Dots come from a fit to the steady state found numerically for the five temperatures presented in panel (a).}
    \label{fig:lambda_2}
\end{figure}

Above the critical temperature
\begin{equation}\label{eq: critical temp}
    T_c=\frac{r_0}{2\mu k_B},
\end{equation}
the steady state in Eq. (\ref{eq:power_law}) is no longer normalizable, creating a delocalization transition. According to \cite{Keidar2025, Roldan2017}, this means that above this temperature, the mean time between resetting events diverges. To our knowledge, this is the first system for which the existence of a NESS under resetting was shown to be temperature-dependent. 
A related transition was identified from a first-passage perspective: below $T_c$, the mean first-passage time of diffusion from the origin to any other point was shown to diverge \cite{Pinsky2020}. The connection between this result and ours is discussed in the End Matter.

In Fig. \ref{fig:lambda_2}(a), we compare the predictions of Eqs. (\ref{eq:power_law}) and (\ref{eq: power law NESS power}) (dashed lines) to numerical solutions of Eq. (\ref{eq: NESS ODE generic}) with $r(x)=(1+x^2)^{-1}$ (solid lines). In Fig. \ref{fig:lambda_2}(b), we compare the observed power-law exponents to the ones predicted by Eq. (\ref{eq: power law NESS power}). Details of the numerical scheme are described in the End Matter. As there is no prediction for the prefactor in Eq. (\ref{eq:power_law}), a constant shift is seen between theory and numerics in Fig. \ref{fig:lambda_2}(a). For visualization purposes, we set the additive constant such that the theoretical asymptotic prediction and the numerical solution will agree at $x=1$. Excellent agreement is found between theoretical predictions and numerics.

\textit{Absence of steady-state for $r(x)$ that decays faster than $\sim1/x^2$.} The fact that for $r_0<2D$ there is no steady-state for resetting rates with $\sim r_0x^{-2}$ tails means that in this case the mean time between resetting events diverges. We stress that this result is not sensitive to the bulk behavior of the resetting rate $r(x)$.

We now consider a resetting rate $r(x)$, which is finite for any finite $x$, and whose tails go to zero faster than $\sim1/x^2$. There thus exists a value $M$ such that $r(x)\leq Dx^{-2}$ for any $x$ whose absolute value is greater than $M$. We will bound $r(x)$ from above by
\begin{equation}
    \bar{r}(x)=\begin{cases}
        r(x)&|x|\leq M,\\
        \frac{D}{x^2}&|x|>M.
    \end{cases}
\end{equation}
By Eq. (\ref{eq: R distribution}), we get that the mean time between resetting events using the resetting strategy $r(x)$ obeys
\begin{equation}
\begin{split}
    \langle R\rangle=&\int_0^\infty \mathbb{E}\left[e^{-\int_0^t r(X_s)\, ds}\right]\,dt\geq\\
    &\int_0^\infty \mathbb{E}\left[e^{-\int_0^t \bar{r}(X_s)\, ds}\right]\,dt=\infty.
\end{split}
\end{equation}
In the last equality, we used the fact that there is no steady-state for a resetting rate that decays as $r_0/x^2$, with $r_0<2D$, which in turn means that the mean time between resetting events diverges (see Eq. (14) in \cite{Keidar2025} and Eq. (30) in \cite{Roldan2017}). For any resetting rate that decays to zero faster than $1/x^2$, a steady state will not emerge. This sets the $\sim x^{-2}$ asymptotics in resetting as the equivalent of the log-potential in equilibrium dynamics.

\textit{Higher dimensions.} We now consider the case of a diffusing particle in $d$ dimensions, with adaptive resetting rate $r(\boldsymbol{x})=r_0\rho^\lambda$ where $\rho=|\boldsymbol{x}|$ is the distance from the origin. Using the $ d$-dimensional Laplacian in spherical coordinates, and utilizing spherical symmetry, we get that the NESS at large $\rho$ obeys
\begin{equation}\label{eq:NESS_ODE_d_dimensions}
    D\frac{\partial^2 p(\boldsymbol{x})}{\partial\rho^2}+\frac{D(d-1)}{\rho}\frac{\partial p(\boldsymbol{x})}{\partial\rho}-r_0\rho^\lambda p(\boldsymbol{x})=0.
\end{equation}

\noindent For $\lambda>-2$, we show in the End Matter that the tails of the NESS are unaffected by dimensionality, and the NESS is given by Eqs. (\ref{eq: stretched exponential}, \ref{eq:streched_exp_paramters}). 

For $\lambda=-2$, we show in the End Matter that the NESS also exhibits power-law tails, but the power depends on the dimension through
\begin{equation}\label{eq:eta_d_dimensions}
    \eta=\frac{1}{2}\left(d-2+\sqrt{\left(d-2\right)^2+\frac{4r_0}{D}}\right).
\end{equation}
This generalizes Eq. (\ref{eq: power law NESS power}), modifying the critical temperature to
\begin{equation}\label{eq:Tc_d_dimension}
    T_c(d)=\frac{r_0}{2d\mu k_B}.
\end{equation}
As can be seen, in higher dimensions the delocalization transition happens at lower temperatures. Finally, we note that the same argument used to establish the absence of a steady state in one dimension for any adaptive resetting rate that decays faster than $x^{-2}$ extends directly to higher dimensions.

\textit{Conclusions.} We have established a sharp localization criterion for diffusion under adaptive resetting. For resetting rates with $r(x)\sim r_0|x|^\lambda$, the asymptotic decay exponent $\lambda=-2$ marks the boundary between localization and delocalization. When $\lambda>-2$, resetting produces a NESS with stretched-exponential tails; when $\lambda<-2$, it is asymptotically too weak to sustain a steady state. At the marginal scaling $r(x)\sim r_0/x^2$, the NESS instead develops temperature-dependent power-law tails and undergoes a finite-temperature delocalization transition. This transition persists in higher dimensions, with a dimension-dependent critical temperature.

These results identify inverse-square resetting as the nonequilibrium counterpart of logarithmic confinement and provide a complete asymptotic classification of localization under power-law adaptive resetting. The analogy is nevertheless fundamentally limited: unlike a Boltzmann distribution, the NESS generally depends nonlocally on the entire resetting-rate profile. Remarkably, its tails are universal, being determined solely by the asymptotic form of the resetting rate.

The delocalization transition associated with logarithmic confinement underlies a broad class of equilibrium critical phenomena. Our results show that an analogous transition can emerge from adaptive resetting, without an underlying confining potential or an equilibrium description. Adaptive resetting thus provides a minimal nonequilibrium setting in which temperature controls not merely the shape of a steady state, but its very existence, and may offer a route toward identifying broader organizing principles for nonequilibrium critical phenomena.

\begin{acknowledgments}
This project has received funding from the European Research Council (ERC) under the European Union’s Horizon 2020 research and innovation program (grant agreement No. 947731).
\end{acknowledgments}

\bibliography{apssamp}

\newpage

\section*{End Matter}

\noindent \textit{Appendix A: Verification of stretched-exponential tails in  Fig. \ref{fig:lambda_great_than_-2}(a, c).}

\begin{figure}[h]
    \centering
    \includegraphics[width=0.85\linewidth]{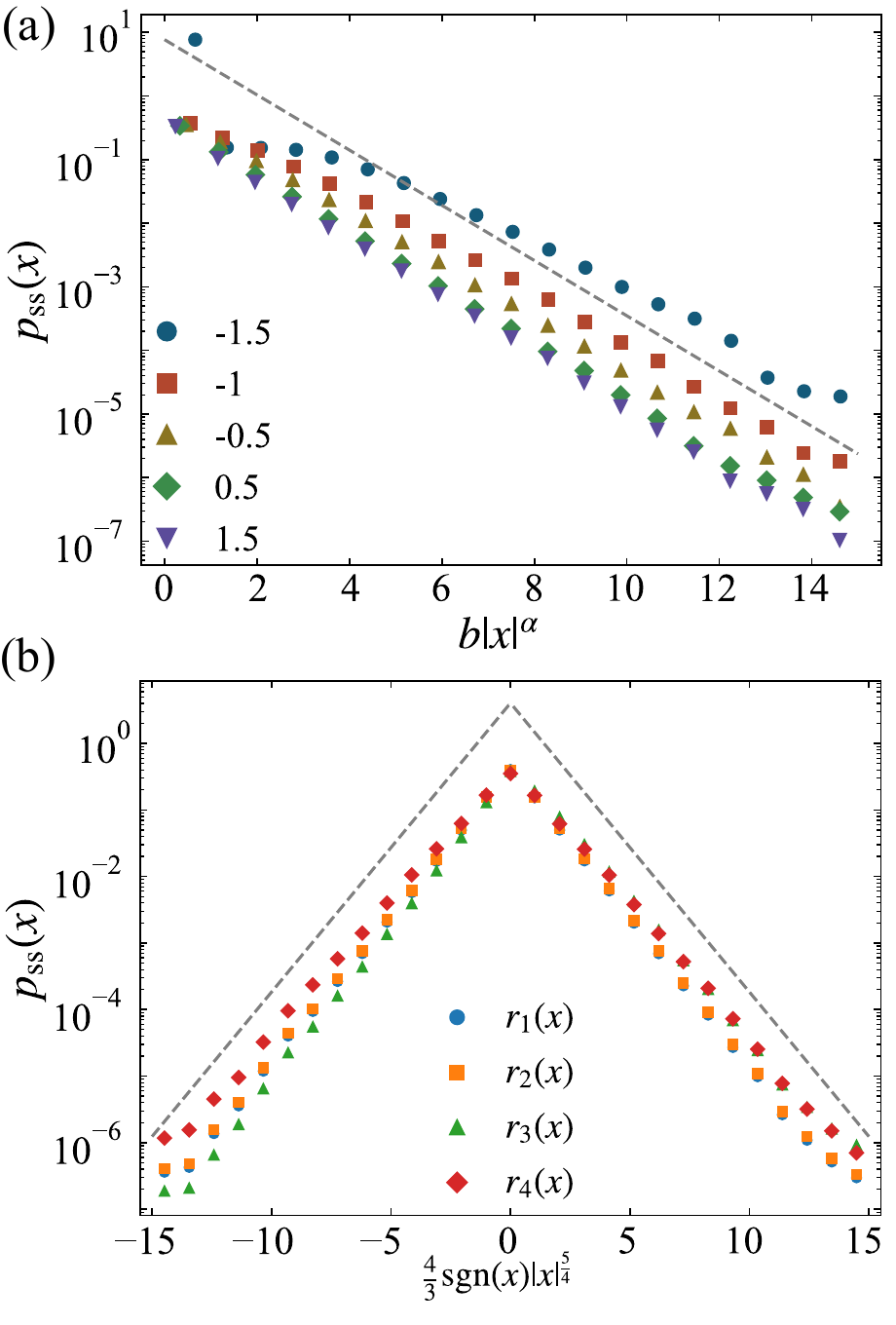}
    \caption{\textbf{a} The NESS presented in Fig. \ref{fig:lambda_great_than_-2}(a). \textbf{b} The NESS presented in Fig. \ref{fig:lambda_great_than_-2}(c). In both cases the y-axis is logarithmic and the x-axis was taken as  $b|x|^\alpha$ where $b$ and $\alpha$ are given by Eq. (\ref{eq:streched_exp_paramters}). All simulation parameters are the same as for Fig. \ref{fig:lambda_great_than_-2}(a, c) except for the trajectory length which was here chosen to be $3\cdot10^4$. Dashed lines with slope $-\mathrm{sgn}(x)$ are added to show the agreement between theory and simulations.}
    \label{fig:scaled_fig}
\end{figure}

\textit{Appendix B: Connection to results regarding the mean first-passage time reported in \cite{Pinsky2020}.}
The paper \cite{Pinsky2020} deals with the mean first-passage time of diffusion under adaptive resetting. That is the mean time $S$ to first reach $x=a$ for a diffusive particle under adaptive resetting. It was shown for diffusion in \cite{Pinsky2020}, and for general process in \cite{Keidar2025}, that the mean first-passage time under adaptive resetting obeys
\begin{equation}
    \langle S\rangle=\frac{\langle\min(S,R)\rangle}{\Pr(S<R)},
\end{equation}
where $R$ is the resetting time. In \cite{Pinsky2020}, it was shown that $\Pr(S<R)>0$ for any continuous resetting rate $r(x)$. Building on that, we can see that
\begin{equation}
    \langle S\rangle=\frac{\langle\min(S,R)\rangle}{\Pr(S<R)}\leq\frac{\langle R\rangle}{\Pr(S<R)}.
\end{equation}

And as it was shown that $\langle S\rangle$ diverges for $T>T_c$, we conclude that in this regime $\langle R\rangle$ diverges as well. For diffusion, a finite mean time between resetting events is a necessary and sufficient condition for a steady state under adaptive resetting \cite{Keidar2025, Roldan2017}. Therefore, we conclude that there should not be a steady state for $T>T_c$, as was shown in the main text using different methods.

\textit{Appendix C: Numerical solution of Eq. (\ref{eq: NESS ODE generic}) for Fig. (\ref{fig:lambda_2}).}
To avoid analyzing long diffusive trajectories to sample the tail in Fig. (\ref{fig:lambda_2}), we instead solved Eq. (\ref{eq: NESS ODE generic}) numerically for $x>0$ with 
\begin{equation}
    r(x)=\frac{1}{1+x^2},
\end{equation}
and boundary conditions $p(0)=1,\,p\left(10^7\right)=0$. After a solution is obtained, we normalize it. To avoid solving this boundary value problem on such a large domain, we instead solved it for a transformed variable $x=\sinh(z)$, for $z\in\left[0,\sinh^{-1}\left(10^7\right)\right]$. In terms of $z$, we arrive at the following equation
\begin{equation}
    T\frac{d^2p(z)}{dz^2}-T\tanh(z)\frac{dp(z)}{dz}-p(z)=0,
\end{equation}
where we set $\mu k_B=1$ and thus $D=T$. This equation is solved using a standard SciPy boundary value problem solver, and the solution is transformed back to the $x$ coordinate.

\textit{Appendix D: Derivations of results for diffusion under power-law adaptive resetting in higher dimensions.}
Starting from Eq. (\ref{eq:NESS_ODE_d_dimensions}) and plugging a stretched exponential ansatz $p(\boldsymbol{x})\propto \exp(-b\rho^\alpha)$, we get
\begin{equation}
    \begin{cases}
        D\alpha b\rho^{\alpha-2}\left(\alpha b\rho^\alpha-\alpha-d+2\right)-r_0\rho^\lambda=0&\text{if $\alpha\neq 1$,}\\
        Db^2-\frac{bD(d-1)}{\rho}-r_0\rho^\lambda=0&\text{if $\alpha=1$.}
    \end{cases}
\end{equation}
Because $\alpha>0$, for large $\rho$ we get
\begin{equation}
D\alpha^2 b^2\rho^{2\alpha-2}-r_0\rho^\lambda=0.
\end{equation}
Arriving at the same result as in Eq. (\ref{eq:streched_exp_paramters}). 

For the case where $\lambda=-2$, we plug a power law ansatz $p(\boldsymbol{x})\propto\rho^{-\eta}$, resulting in
\begin{equation}
    \eta^2+(2-d)\eta-\frac{r_0}{D}=0.
\end{equation}
The solution for $\eta$ is given by Eq. (\ref{eq:eta_d_dimensions}). For this solution to be normalizable, we must require that $\eta>d$, which gives the critical temperature in Eq. (\ref{eq:Tc_d_dimension}).

\end{document}